\documentclass[12pt,pre,aps,showpacs,preprint]{revtex4}

\usepackage{graphicx}
\usepackage{amsfonts}

\begin{document}

\title{New Duality Relations for Classical Ground States}

\author{S. Torquato}


\affiliation{\emph{School of Natural Sciences}, \emph{Institute for Advanced Study}, Princeton
NJ 08540}

\affiliation{\emph{Department of Chemistry}, \emph{Princeton University}, Princeton
NJ 08544}

\affiliation{\emph{Princeton Institute for the Science and Technology of Materials, Princeton University}, Princeton NJ 08544}

\affiliation{\emph{Program in Applied and Computational Mathematics}, \emph{Princeton
University}, Princeton NJ 08544}

\affiliation{\emph{Princeton Center for Theoretical Physics, Princeton University}, Princeton NJ 08544}

\author{F. H. Stillinger}


\affiliation{\emph{Department of Chemistry}, \emph{Princeton University}, Princeton
NJ 08544}

\begin{abstract}

We derive new duality relations that link the energy of configurations associated with a 
class of soft pair potentials to the corresponding energy of the dual (Fourier-transformed) potential.
We apply them by showing how information
about the classical ground states of short-ranged potentials 
can be used to draw new conclusions about the nature
of the ground states of long-ranged potentials and vice versa.
They also lead to  bounds on the $T=0$ system energies  in density intervals
of phase coexistence, the identification of a one-dimensional system that exhibits
an infinite number of ``phase transitions," and  a conjecture regarding the ground states 
of purely repulsive monotonic potentials.

\end{abstract}

\maketitle

 While classical ground states are readily produced
by slowly freezing liquids in experiments and computer simulations, our theoretical
understanding of them is far from complete. Much of the 
progress to rigorously identify ground states for given interactions has been for
lattice models, primarily in one dimension \cite{Ra87}. The solutions
in $d$-dimensional Euclidean space $\mathbb{R}^d$ for $d \ge 2$
are considerably more challenging. Recently, a ``collective-coordinate" approach has been
used to study and ascertain ground states in $\mathbb{R}^2$ and $\mathbb{R}^3$ 
for a  class of interactions \cite{Uc04,Su05}. A surprising conclusion
of Ref. \cite{Uc04} is that there exist nontrivial {\it disordered} ground
states without any long-range order \cite{footnote},  in addition to the expected periodic ones.
Despite these advances,  new theoretical tools are required to make further progress.

Here we derive new duality relations for a  class of soft pair potentials 
that can be applied to classical ground states.
We consider soft interactions since they are easier to treat
theoretically and possess great importance in soft-matter systems,
such as colloids, microemulsions,  and polymers \cite{La00,Ml06,Ka07}.
The duality relations link the energy of configurations for a
pair potential $v(r)$ to that for  the dual (Fourier-transformed) potential.
Applications of the duality relations lead to some novel results.

For a configuration ${\bf r}^N\equiv {\bf r}_1,{\bf r}_2,\ldots,{\bf r}_N$ of $N \gg 1$ particles in volume $V\subset \mathbb{R}^d$ with 
stable  pairwise interactions \cite{footnote1},
$U({\bf r}^N)=\frac{1}{N}\sum_{i=1, j=1} v(r_{ij})$ is twice the total potential energy per particle [plus the ``self-energy" $v(0)$], 
where $v(r)$ is a radial pair potential function and $r_{ij}=|{\bf r}_j-{\bf r}_i|$.
A {\it classical ground-state} configuration  is one that minimizes $U({\bf r}^N)$.
Since we allow for disordered ground states, we consider
the general ensemble setting.
The {\it ensemble average} of $U$ for a statistically homogeneous and isotropic system in the thermodynamic
limit is given by 
\begin{equation}
\langle U({\bf r}^N) \rangle= v(r=0)+\rho \int_{\mathbb{R}^d} v(r) g_2(r) d{\bf r},
\label{g2}
\end{equation}
where $\rho=\lim_{N\rightarrow \infty,V\rightarrow \infty}N/V$ is the number density
and $g_2(r)$ is the {\it pair correlation function}. It is crucial to introduce
the {\it total correlation function} $h(r)\equiv g_2(r)-1$, which decays
to zero for a disordered system. We consider
those stable radial pair potentials $v(r)$ that are bounded and absolutely integrable
and call such functions {\it admissible}. Thus,
the corresponding  Fourier transform ${\tilde v}(k)$ in $d$ dimensions \cite{To03} 
at wavenumber $k$ exists, which we also take to be admissible, and
\begin{equation}
\langle U({\bf r}^N) \rangle= v(r=0)+\rho{\tilde v}(k=0)+\rho \int_{\mathbb{R}^d} v(r) h(r) d{\bf r}.
\label{h}
\end{equation}

\noindent{\bf Lemma.} For {\it any ergodic configuration} in $\mathbb{R}^d$, 
the following duality relation holds:
\begin{equation}
\int_{\mathbb{R}^d} v(r) h(r) d{\bf r}= \frac{1}{(2\pi)^d}\int_{\mathbb{R}^d} {\tilde v(k)} {\tilde h}(k) d{\bf k}
\label{plancherel}
\end{equation}
If such a configuration is a ground state, then left and right
sides of (\ref{plancherel}) are {\it minimized}.

\noindent{\bf Proof:} Identity (\ref{plancherel}) follows from Parseval's theorem,
assuming that ${\tilde h}(k)$ or the structure factor $S(k)\equiv 1+\rho{\tilde h}(k)$ exists. 
From  (\ref{h}) and (\ref{plancherel}), we see that
the both sides of (\ref{plancherel}) are minimized for any ground-state structure,
although the duality relation applies to general structures \cite{noticed}.

\noindent{\bf Remark:} 

\noindent{Whereas $h(r)$ always
characterizes a point pattern, its Fourier transform
${\tilde h}(k)$  is generally not the total correlation function of a point pattern
in reciprocal space.
It is when  $h(r)$ characterizes a Bravais lattice $\Lambda$ \cite{lattice}
that  ${\tilde h}(k)$ is the total correlation function of a point pattern, namely,
the reciprocal Bravais lattice ${\tilde \Lambda}$.}
\smallskip

\smallskip

\noindent{\bf Theorem 1.} 
If an admissible pair potential $v(r)$ has a Bravais lattice $\Lambda$ 
ground-state structure at number density $\rho$, then we have the following duality relation 
for the minimum $U_{min}$ of $U$:
\begin{equation}
v(r=0)+ {\sum_{{\bf r} \in \Lambda}}^{\prime} v(r) = \rho {\tilde v}(k=0)+ 
\rho {\sum_{{\bf k} \in {\tilde \Lambda}}}^{\prime} {\tilde v}(k),
\label{duality}
\end{equation}
where the prime on the sum denotes the zero vector should be omitted, ${\tilde \Lambda}$ denotes the reciprocal Bravais lattice \cite{footnote3},
and ${\tilde v}(k)$ is the dual pair potential, which automatically satisfies the 
stability condition, and therefore is admissible.
Moreover, the minimum $U_{min}$ of $U$
for any ground-state structure of the dual potential ${\tilde v}(k)$, is
bounded from above by the corresponding real-space {\it minimized} quantity $U_{min}$ 
or, equivalently, the right side of (\ref{duality}), i.e.,
\begin{equation}
{\tilde U}_{min} \le U_{min}=\rho {\tilde v}(k=0)+ \rho  {\sum_{{\bf k} \in {\tilde \Lambda}}}^{\prime} {\tilde v}(k).
\label{bound}
\end{equation}
Whenever the reciprocal lattice  ${\tilde \Lambda}$ at {\it reciprocal lattice density} 
${\tilde \rho}=\rho^{-1}(2\pi)^{-d}$ is a ground state of ${\tilde v}(k)$,
the inequality in  (\ref{bound}) becomes an equality.
On the other hand, if an admissible dual potential ${\tilde v}(k)$ has
a Bravais lattice ${\tilde \Lambda}$ at number density ${\tilde \rho}$,
then
\begin{equation}
U_{min} \le {\tilde U}_{min}={\tilde \rho} v(r=0)+ {\tilde \rho}  
{\sum_{{\bf r} \in {\Lambda}}}^{\prime} v(r),
\label{bound2}
\end{equation}
where equality is achieved when the real-space ground state is the lattice $\Lambda$ 
reciprocal to ${\tilde \Lambda}$.

\noindent{\bf Proof:} 
The radially averaged total correlation function
for a Bravais lattice, which we now assume to be a ground-state structure, is given by
$h(r)=\frac{1}{\rho s_1(r)}\sum_{n=1} Z_n \delta(r-r_n) -1$,
where $s_1(r)$ is the surface area of a $d$-dimensional sphere of radius $r$, 
$Z_n$ is the coordination number (number of points) at the radial distance
$r_n$, and $\delta(r)$ is a radial Dirac delta function. Substitution of this expression
and the corresponding one for ${\tilde h}(k)$ into  (\ref{plancherel}) yields
$v(r=0)+ \sum_{n=1} Z_n v(r_n) = \rho {\tilde v}(k=0)+ \rho \sum_{n=1} {\tilde Z}_n{\tilde v}(k_n)$,
where ${\tilde Z}_n$ is the coordination number in the reciprocal lattice at the radial distance $k_n$.
Recognizing that $\sum_{n=1} Z_n v(r_n)= \sum_{{\bf r} \in \Lambda}^{\prime} v(r)$ (leading to $U_{min}$)
and $\sum_{n=1} {\tilde Z}_n {\tilde v}(k_n)= \sum_{{\bf k} \in {\tilde \Lambda}}^{\prime} {\tilde v}(k)$ 
yields the duality relation (\ref{duality}).
However, there may be non-Bravais lattice structures \cite{lattice} that have lower energy
than the reciprocal lattice so that ${\tilde U}_{min}\le U_{min}$ \cite{To07}. 
Inequality (\ref{bound2}) follows in the same manner as (\ref{bound})
when the ground state of ${\tilde v}(k)$ is known to
be a Bravais lattice.

\noindent{\bf Remarks:}

\noindent{1. Whenever equality in relation (\ref{bound}) is achieved,
then a ground state structure of the dual potential ${\tilde v}(k=r)$
evaluated at the real-space variable $r$ is the Bravais lattice ${\tilde \Lambda}$
at density ${\tilde \rho}=\rho^{-1}(2\pi)^{-d}$.}
\smallskip                                                            

\noindent{2. The zero-vector contributions on both sides of the duality relation (\ref{duality})
are crucial in order to establish a relationship between the real-
and reciprocal-space ``lattice" sums indicated therein \cite{yukawa}.}

\smallskip

\noindent{3. We identify below specific instances in which
the strict inequalities in (\ref{bound}) and (\ref{bound2})
apply, including a theorem and a one-dimensional system with unusual properties.}
\smallskip

\noindent{\bf Theorem 2.} 
Suppose that for admissible potentials there exists a range of densities over which
the ground states are side by side coexistence
of two distinct structures whose parentage are two 
different Bravais lattices, then the strict inequalities in
(\ref{bound}) and (\ref{bound2}) apply at any density
in this density-coexistence interval.

\noindent{\bf Proof:} This follows immediately from the Maxwell double-tangent construction
in the $U$-$\rho^{-1}$ plane,
which ensures that $U$ in the coexistence
region at density $\rho$ is lower than either of the 
two Bravais lattices.

As we will see, the duality relations of Theorem 1 
will enable one to use information about ground states of short-ranged potentials
to draw new conclusions about the nature
of the ground states of long-ranged potentials and vice versa.
Moreover, inequalities (\ref{bound}) and (\ref{bound2}) provide a computational tool
to estimate ground-state energies or eliminate
candidate ground-state structures as obtained from annealing simulations.
We will now examine the ground states of several classes
of admissible functions.

{\it Admissible functions with compact support.--} Recently, the ground states
of a  class of oscillating real-space potentials $v(r)$ as defined
by the family of Fourier transforms  with compact support
such that ${\tilde v}(k)$ is positive for $0 \le k < K$
and zero otherwise have been studied \cite{Uc04,Su05}. Clearly, ${\tilde v}(k)$
is admissible.  S{\" u}t{\^ o} \cite{Su05} showed
that in $\mathbb{R}^3$ the corresponding  real-space potential $v(r)$,
which oscillates about zero,  has the body-centered
cubic (bcc) lattice as its unique ground state at the real-space density
$\rho=1/(8\sqrt{2}\pi^3)$ (with $K=1$). Moreover, he 
showed that for densities greater than $1/(8\sqrt{2}\pi^3)$,
the ground states are degenerate such that the face-centered cubic (fcc),
simple hexagonal (sh), and simple cubic (sc) lattices are ground states
at and above the respective densities $1/(6\sqrt{3}\pi^3)$, $\sqrt{3}/(16\sqrt{2}\pi^3)$, and
$1/(8\sqrt{2}\pi^3)$.

\begin{figure}[bthp]
\centerline{\includegraphics[height=2.2in,keepaspectratio,clip=]{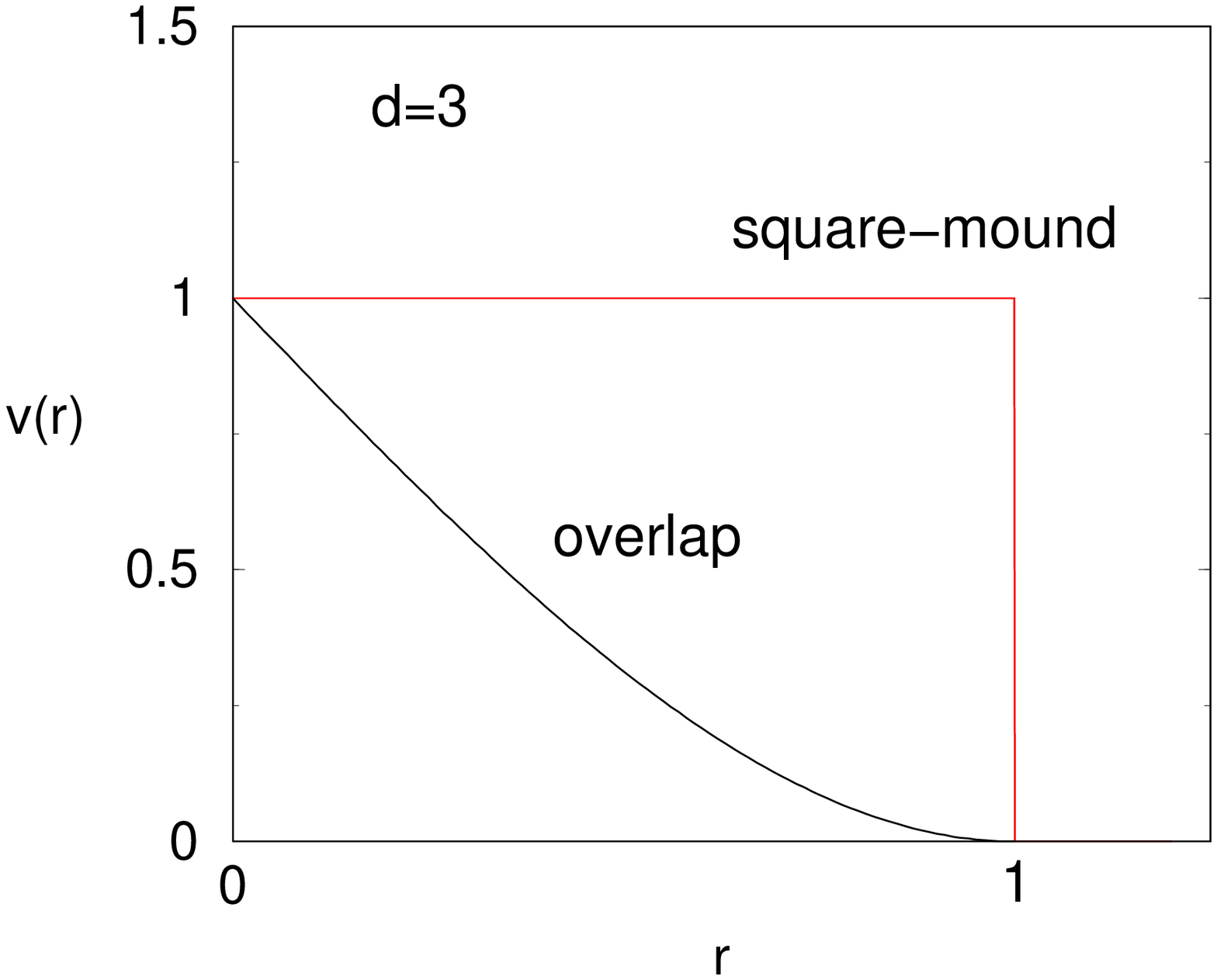}\hspace{0.25in}
\includegraphics[height=2.1in,keepaspectratio,clip=]{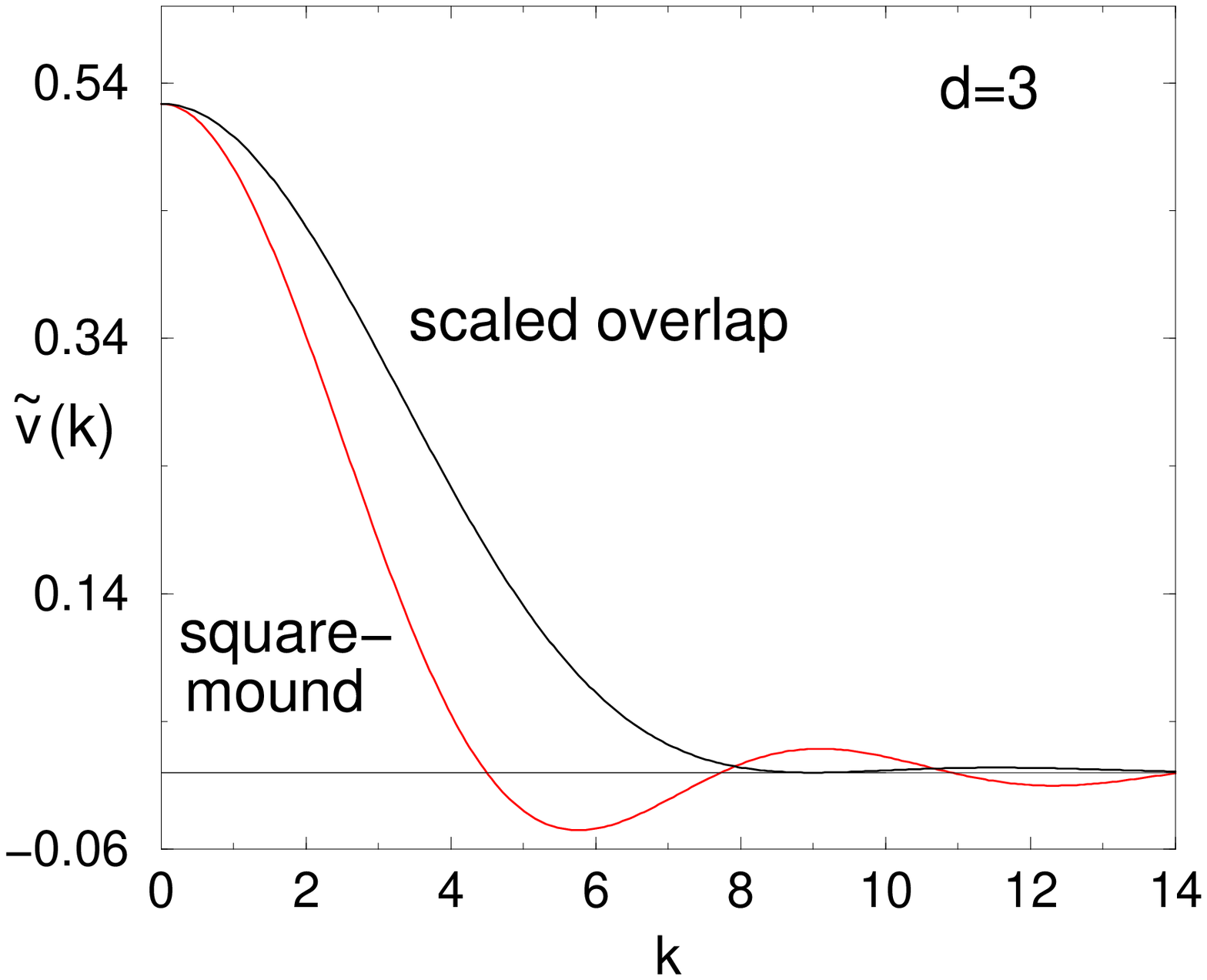}}
\caption{ Left: The localized square-mound potential
[$v(r)=\epsilon=1$ for $0 \le r <1$ and zero otherwise] and overlap potential [$v(r)=1-3r/2+r^3/2$ for $0 \le r <1$ and zero otherwise]
in $\mathbb{R}^3$.
Right: The  delocalized dual square-mound potential ${\tilde v}(k)=\pi^{3/2}J_{3/2}(k)/(2k)^{3/2}$
multiplied by $\pi^3/6$ and dual overlap potential ${\tilde v}(k)=6\pi^2 J_{3/2}(k/2)/k^3$.}
\label{compact}
\end{figure}

The long-range behavior of the real-space
oscillating potential $v(r)$ might be regarded to be unrealistic by some. However, since
all of the aforementioned ground states are Bravais
lattices, the duality relation (\ref{duality}) can be applied
here to infer the ground states of real-space potentials
with compact support. Specifically, application of the duality theorem in $\mathbb{R}^3$
and S{\" u}t{\^ o}'s results enables us to conclude that for the real-space potential
$v(r)$ that is positive for $0 \le r < D$ and zero otherwise,
the fcc lattice (dual of the bcc lattice) is a ground state
at the density $\sqrt{2}$ and the ground states are degenerate such that the bcc, sh
and sc lattices are ground states at and below the respective densities $(3\sqrt{3})/4$, $2/\sqrt{3}$, and
$1$ (taking $D=1$). Specific examples
of such real-space potentials, for which the ground
states are not rigorously known, include the ``square-mound" potential \cite{Hi57}
[$v(r)=\epsilon >0$ for $0 \le r <1$ and zero otherwise] and 
the ``overlap" potential \cite{To03}, equal to the intersection
volume of two $d$-dimensional spheres of diameter $D$ whose
centers are separated by a distance $r$ divided by the
volume of a sphere, and thus has support in the interval
$[0,D)$ \cite{number}. The $d$-dimensional Fourier transforms
of the square mound  and overlap potentials are 
$\epsilon 2^{d/2}J_{d/2}(k)/(k\pi)^{d/2}$
and $2^d\pi^{d/2}\Gamma(1+d/2)J^2_{d/2}(k/2)/k^d$, respectively, with $D=1$.
Figure \ref{compact} shows the real-space and dual potentials
for these examples in $\mathbb{R}^3$. The densities at which the aforementioned lattices
are ground states  are easily understood by appealing
to either the square-mound or overlap potential. The fcc lattice is a ground state
at the density $\sqrt{2}$ because at this value, where the nearest-neighbor (NN) distance
is unity, and lower densities, the energy
is zero. At a slightly higher density, each of the 12 nearest neighbors
contribiutes an amount $\epsilon$ to the lattice energy.  At densities lower than $\sqrt{2}$,
there  are an uncountably infinite number 
of degenerate ground states. This includes the bcc, sh
and sc lattices, which join in as ground states
at and below the respective densities $(3\sqrt{3})/4$, $2/\sqrt{3}$, and
$1$ because those are the threshold values at which these structures
have lattice energies that change discontinuously from some positive value (determined
by nearest neighbors only) to zero. Moreover, any structure, periodic or not,
in which the NN distance is greater than unity is
a ground state. 

However, at densities corresponding to NN distances
that are less than unity, determination of the possible ground-state structures is considerably
more difficult. For example, it has been argued in Ref. \cite{Ml06} (with good reason) that real-space
potentials whose Fourier transforms oscillate about zero will
exhibit polymorphic crystal phases in which the particles
that comprise a cluster sit on top of each other. The square-mound potential
is a special case of this class of potentials and the fact
that it is a simple piecewise constant function
allows for a rigorous analysis of the clustered ground states
for densities in which the NN distances
are less than the distance at which the discontinuity
in $v(r)$ occurs \cite{To07}.

\begin{figure}[bthp]
\centerline{\includegraphics[height=2.2in,keepaspectratio,clip=]{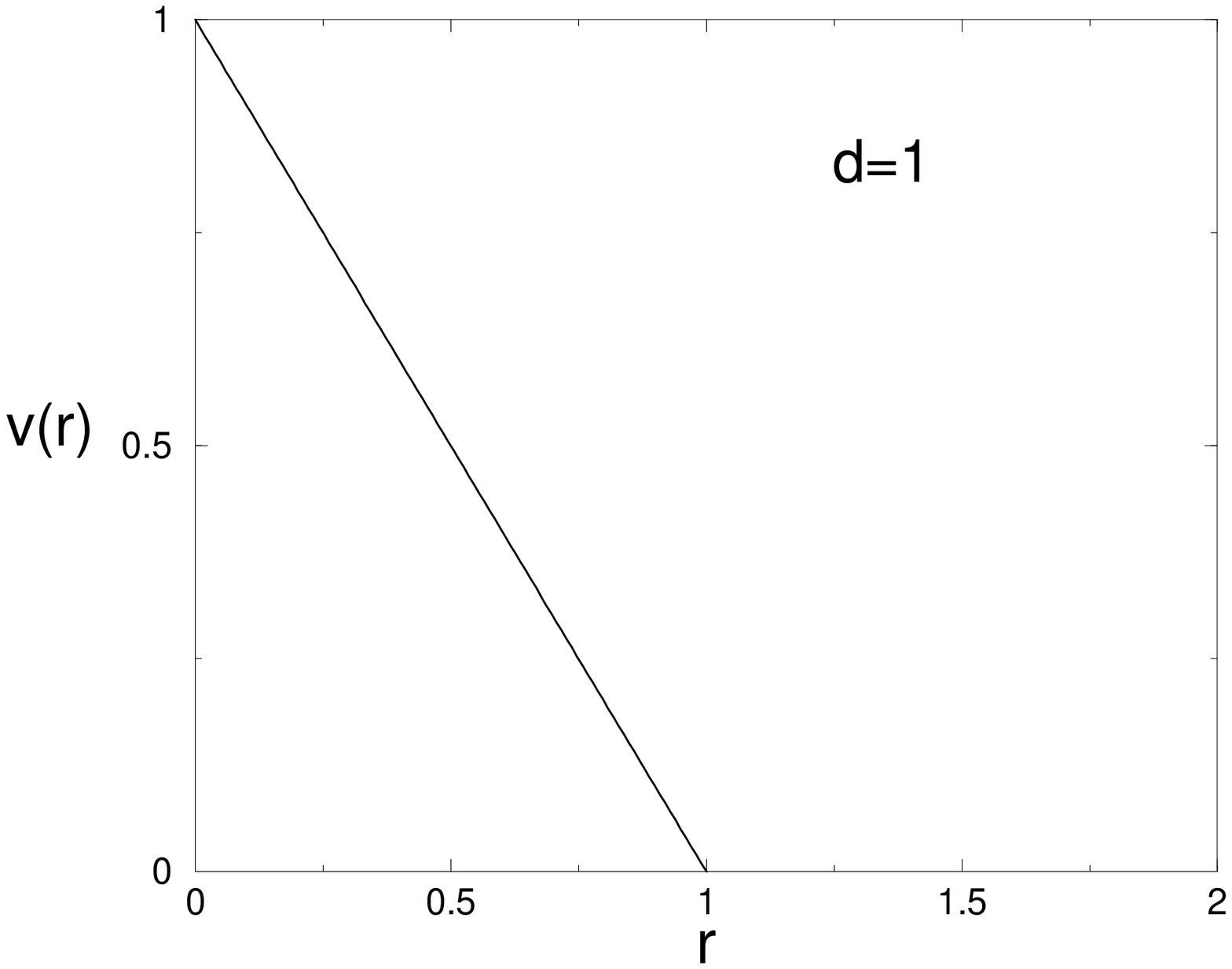}\hspace{0.2in}
\includegraphics[height=2.2in,keepaspectratio,clip=]{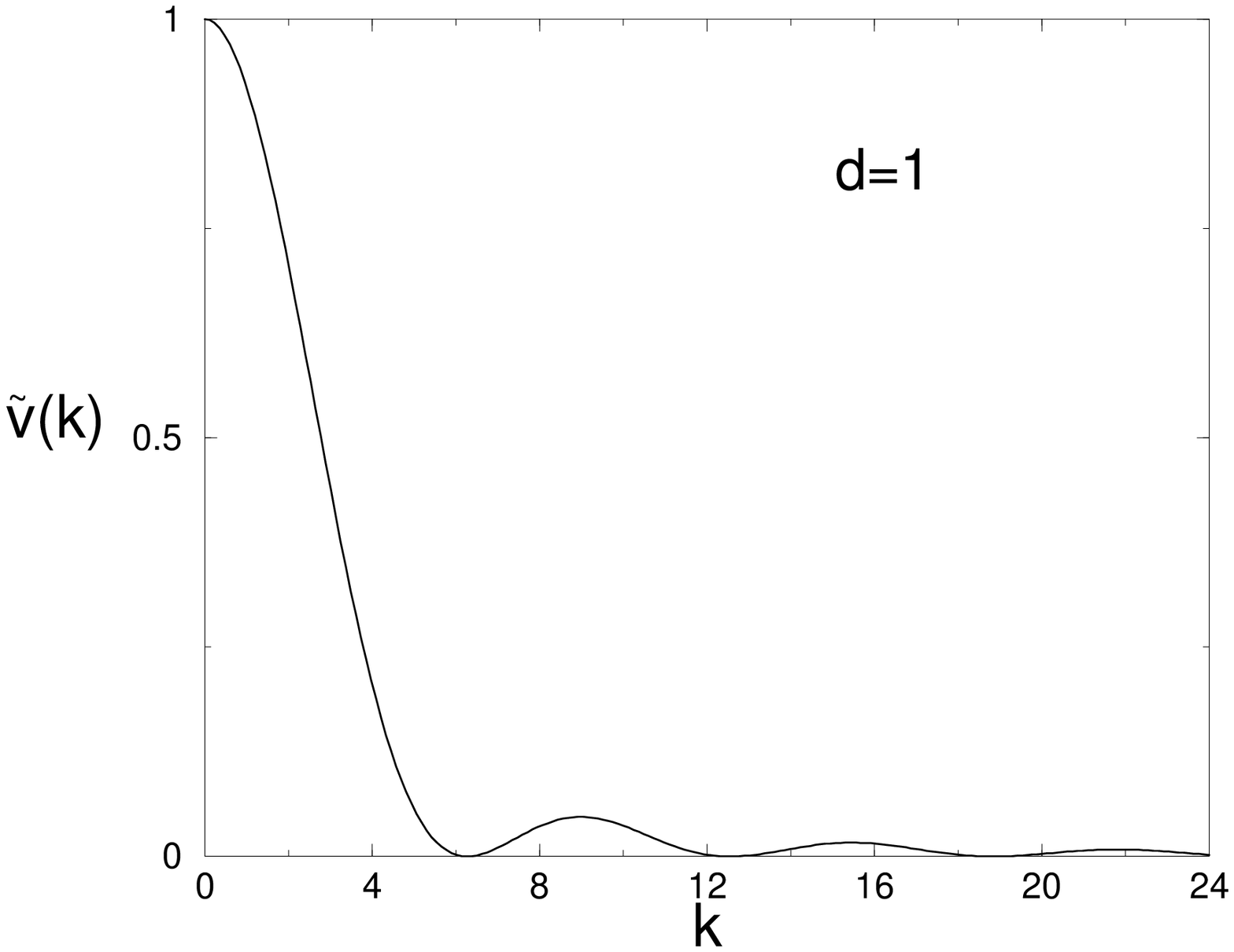}}
\caption{ Left: The ``localized" overlap potential in $\mathbb{R}$:
$v(r)=1-r$ for $0 \le r \le 1$ and zero otherwise.
Right: The corresponding ``delocalized" dual potential: ${\tilde v}(k)= 4 \sin^2(k/2)/k^2$.}
\label{linear}
\end{figure}

{\it Nonnegative admissible functions.--} Another interesting class of admissible functions
are those in which both $v(r)$ and ${\tilde v}(k)$ are nonnegative (i.e., purely repulsive) for their entire domains.
The ``overlap" potential discussed above is an example. Here we show that
the dual ${\tilde v}(k)= 4 \sin^2(kD/2)/(kD)^2$ of the overlap potential for
$d=1$ [$v(r)=1-r/D$ for $0 \le r \le D$ and zero otherwise] (see Fig. \ref{linear}) exhibits rich behavior. 
For any density $\rho$, it can be shown that the unique Bravais (integer) lattice
with spacing $\rho^{-1}$ is a ground-state structure \cite{To03,To07}. Moreover,
using Theorem 1,  we can show that for any $\rho=m$, where $m$ is a positive integer,
the integer lattice at reciprocal density ${\tilde \rho}=(2\pi m)^{-1}$
is a ground-state structure for for the dual potential ${\tilde v}(k)$; however,
at non-integer density $\rho$, ground-state structures for ${\tilde v}(k)$
are generally non-Bravais lattices, establishing the strict
inequality of duality relation (\ref{bound}) \cite{To07}. This
latter result implies that for ${\tilde v}(k)$, the system
undergoes an infinite number of ``phase transitions" from
Bravais to non-Bravais lattices over the entire
density range. This one-dimensional example is
interesting in its own right and further details 
about its ground states will be given elsewhere \cite{To07}.

Another interesting example of nonnegative admissible functions is
the Gaussian (core) potential $v(r)=\epsilon \exp[-(r/\sigma)^2]$ \cite{St76}, which has been used
to model interactions in polymers \cite{La00}. The 
dual potentials are self-similar Gaussian functions for any $d$. 
The potential function pairs for the case $d=3$ with $\epsilon=1$ and $\sigma=1$ are $v(r)=\exp(-r^2)$
and ${\tilde v}(k)=\pi^{3/2}\exp(-k^2/4)$. Simulations \cite {St76} indicate that
at sufficiently low densities in $\mathbb{R}^3$, the fcc lattices are the ground state structures
for $v(r)$.
For the range $0 \le \rho < \pi^{-3/2}$, fcc is favored over bcc \cite{St79}.
If equality in (\ref{bound}) is achieved for this density range, the duality
theorem would imply that the bcc lattices in the range $ (4\pi)^{-3/2} \le {\tilde \rho} < \infty$
are the ground state structures for the dual potential. 
Previous work \cite{St76} has verified that this
is the case, except in a narrow density interval
of fcc-bcc coexistence $0.17941 \le \rho \le 0.17977$ around  $\rho=\pi^{-3/2}\approx 0.17959$. In the coexistence interval, however, Theorem 2
states the strict inequalities in (\ref{bound}) and (\ref{bound2}) must apply. Importantly, the 
ground states here are not only non-Bravais lattices, they are not even periodic \cite{coexistence}. 
In $\mathbb{R}^2$, the triangular lattices apparently are the ground states
for the Gaussian  potential at all densities (but  there is
no proof), and therefore would not exhibit a phase transition.
Proposition 9.6 of Ref. \cite{Co07} enables us to conclude 
that the integer lattices are the ground states of the Gaussian
potential for all densities in $\mathbb{R}$.

{\it Completely monotonic (CM) admissible functions.--} A radial function $f(r)$ is completely
monotonic if it possesses derivatives $f^{(n)}(r)$ for all $n\ge 0$ and if 
$(-1)^n f^{(n)}(r) \ge 0$. Not all CM
functions are admissible (e.g., the power-law potential
$1/r^{\gamma}$ in $\mathbb{R}^d$ is inadmissible). Examples of admissible ones
in $\mathbb{R}^d$ include $\exp(-\alpha r)$ for $\alpha > 0$ and $1/(r+\alpha)^\beta$
for $\alpha > 0$, $\beta > d$.

Remarkably, the  ground states of the pure exponential potential have
not been studied. Here we apply the duality relations to the real-space potential $v(r)=\exp(-r)$ in $\mathbb{R}^d$ and
its dual  ${\tilde v}(k)=c(d)/(1+k^2)^{(d+1)/2}$ [where
$c(d)=2^d \pi^{(d-1)/2} \Gamma((d+1)/2)$], which has a slow power-law
decay of $1/k^{d+1}$ for large $k$. Note that ${\tilde v}(k)$ 
is a CM admissible function in $k^2$, and both $v(r)$ and ${\tilde v}(k)$
are nonnegative admissible functions. We have evaluated   lattice sums
for the exponential potential for a variety of Bravais and periodic structures
in $\mathbb{R}^2$ and $\mathbb{R}^3$. In $\mathbb{R}^2$, we found
that the triangular lattices are favored for all $\rho$ (as
in the Gaussian case).
If equality in (\ref{bound})
is achieved, the triangular lattices are also the ground states for the 
slowly decaying dual potential
${\tilde v}(k)=2\pi/(1+k^2)^{3/2}$ for all $\rho$.
In $\mathbb{R}^3$,  the fcc lattices are favored at low densities
($0 \le \rho \le 0.017470$) and bcc lattices are favored at high densities 
($ 0.017470 \le \rho < \infty$). The Maxwell double-tangent construction
reveals that there is a very narrow density interval $0.017469 \le \rho \le 0.017471$
of fcc-bcc coexistence. The exponential potential
appears to behave qualitatively like the Gaussian. If equality in (\ref{bound})
applies outside the coexistence interval, Theorem 1 would predict 
that the ground states of the  dual potential ${\tilde v}(k)=8\pi/(1+k^2)^2$ 
are the fcc lattices for $0 \le {\tilde \rho} \le 0.230750$
and the bcc lattice for $0.230777 \le {\tilde \rho} < \infty$ \cite{henry1}.

\noindent{\bf Conjecture.} 
The Gaussian  potential, exponential
potential, the dual of the exponential potential, and any other admissible
 potential function that is completely monotonic (CM) in distance or squared distance share the same
ground-state structures in $\mathbb{R}^d$ for $2 \le d \le 8$ and $d=24$,
albeit not at the same densities. 
For any such potential
function, the ground states are the Bravais lattices corresponding to the
densest known sphere packings \cite{packing} for $0 \le \rho \le \rho_1$ and the reciprocal
Bravais lattices for $\rho_2 \le \rho <\infty$, where $\rho_1$
and $\rho_2$ are the density limits of phase coexistence
of the low- and high-density phases, respectively. Whenever,
the Bravais and reciprocal lattices are self-dual ($d=2,4, 8$ and $24$)
$\rho_1=\rho_2$, otherwise $\rho_2 > \rho_1$ (which occurs for $d=3,5,6$ and 7).

This conjecture is bolstered by the recent work of Cohn and Kumar \cite{Co07}, 
who rigorously proved that certain configurations of points interacting
with CM potentials on the surface of the unit sphere 
in arbitrary dimension were energy-minimizing \cite{henry2}.  

In summary, we have derived and applied duality relations
to help quantify and  identify ground states 
for pair potentials that arise in soft-matter systems.
Elsewhere, we will apply the duality relations to
a broader category of functions beyond the pure Gaussians that are self-similar 
under Fourier transformation and will show
that our formalism be extended to obtain
corresponding duality relations for potential
functions that also include three-body and higher-order
interactions \cite{To07}.

\begin{acknowledgments}
The authors thank H. Cohn for helpful discussions.
S. T. thanks  the Institute for Advanced Study 
for their hospitality during his  stay there.
This work was supported by the DOE
under Grant No. DE-FG02-04ER46108.
\end{acknowledgments}


\begin{thebibliography}{10}
\bibitem{footnote}
Note that the classical hard-sphere system exhibits
disordered classical ground states (even if trivial) from zero density up to the
``freezing point" because at any instant in time the total
system energy is at its minimum of zero.
\bibitem{noticed}                                                                                       The general duality relation (\ref{plancherel}) does not seem
to have been noticed or exploited before, although it was used 
for a specific pair interaction in Ref. \cite{To03}. 
On account of the ``uncertainty principle" for Fourier pairs,
it provides an efficient way to compute energies per particle
for a non-localized (long-ranged) potential, say $v(r)$, by evaluating
the equivalent integral in reciprocal space for the corresponding
localized (compact) dual potential ${\tilde v}(k)$.
\bibitem{footnote3}
Note that the duality relation (\ref{duality}) is an expanded form 
of the well-known Poisson summation formula, i.e., the zero-vector
contributions are explicitly displayed.
\bibitem{yukawa} To emphasize this point, Ref. \cite{To07} discusses
why the Yukawa (screened-Coloumb) potential  is inadmissible.
\bibitem{coexistence}
Importantly, the
ground states here are not only non-Bravais lattices, they are not even periodic.
The ground  states are side-by-side coexistence of two crystal domains
whose shapes and relative orientations are complicated functions of $\rho$.
\bibitem{henry1}
In $\mathbb{R}$, it  follows from Ref. \cite{Co07} that since the integer lattices
are the ground states of the Gaussian potential, they
are the ground states of both the exponential
potential and its dual evaluated at $k=r$ [i.e., $v(r)=2/(1+r^2)$].
\bibitem{henry2}
If the ground states for CM functions of squared distance in $\mathbb{R}^d$ (the Gaussian
function being a special case) can be proved for any $d\ge 2$, it follows from Ref. \cite{Co07} that the
CM functions of distance share the same ground states.
\end{thebibliography}
\end{document}